\documentclass{article}
\usepackage{spconf,amsmath,amssymb,graphicx,hyperref}
\usepackage{xcolor}
\usepackage{booktabs}
\usepackage{multirow}
\usepackage{tikz}
\usetikzlibrary{arrows.meta,positioning}

\title{MIND THE ACCENT GAP: BRITISH ACCENT ROBUSTNESS \\IN SPEECH-DRIVEN FINANCIAL VOICE ASSISTANTS}
\makeatletter
\gdef\@name{%
  {\em Aadam Haq, Oggi Rudovic, Malcolm Chadwick, Jay Rainey, Shucong Zhang, Ricardo Guerrero}\\
  {\em Sourav Bhattacharya and Maja Pantic}\\
}
\makeatother
\address{NatWest, UK}

\begin{document}
\maketitle

\begin{abstract}

AI voice assistants often use Automatic Speech Recognition (ASR) with an LLM reasoning, yet, existing systems struggle with regional British accents (Scottish, Irish, Welsh), since most ASR models are trained predominantly on American English voice data. Consequently, errors carry through to the LLM stage, corrupting tool-call arguments and producing wrong or missing responses, which is especially costly in finance. Deployable ASR must also meet tight latency and memory budgets, making an accent-robust model choice even harder. We introduce CavaBench, the first internally collected benchmark of spoken financial queries, and use it to evaluate a range of ASR models and their full ASR-LLM pipeline behavior across self-reported British accents. We find that WER strongly predicts downstream tool-calling accuracy (r = -0.93) but can fail to reflect task-level performance, with accent-related failures varying substantially across models and acoustic conditions. These findings guide the design of more inclusive, reliable voice-based financial assistants.
\end{abstract}


\section{Introduction}
\label{sec:intro}

Most finance virtual assistants are text-based~\cite{cfpb2023chatbots}; speech interfaces are a more natural, engaging alternative~\cite{rzepka2022voice} but add Automatic Speech Recognition (ASR) as a source of error. State-of-the-art voice assistants (VAs) increasingly combine ASR with Large Language Model (LLM) reasoning in a cascade. These systems generalize well for well-represented language groups but remain fragile to regional accents, especially British English accents (Scottish, Irish, Welsh), because most ASR systems are optimized for American English. Accent-induced errors compound downstream, so VAs misrepresent user intent and systematically underperform for specific accent groups.

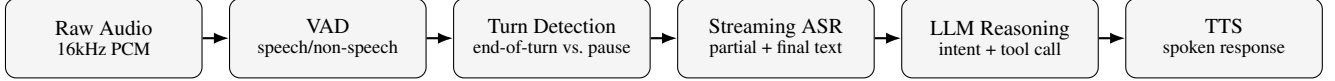
\begin{figure*}[t]
\centering
\begin{tikzpicture}[
  box/.style={draw, rounded corners, minimum height=1.05cm, minimum width=2.6cm, align=center, font=\footnotesize, fill=gray!8},
  arr/.style={-{Latex[length=2mm]}, thick}
]
\node[box] (audio) {Raw Audio\\[-2pt]\scriptsize 16kHz PCM};
\node[box, right=0.35cm of audio] (vad) {VAD\\[-2pt]\scriptsize speech/non-speech};
\node[box, right=0.35cm of vad] (turn) {Turn Detection\\[-2pt]\scriptsize end-of-turn vs.\ pause};
\node[box, right=0.35cm of turn] (asr) {Streaming ASR\\[-2pt]\scriptsize partial + final text};
\node[box, right=0.35cm of asr] (llm) {LLM Reasoning\\[-2pt]\scriptsize intent + tool call};
\node[box, right=0.35cm of llm] (tts) {TTS\\[-2pt]\scriptsize spoken response};

\draw[arr] (audio) -- (vad);
\draw[arr] (vad) -- (turn);
\draw[arr] (turn) -- (asr);
\draw[arr] (asr) -- (llm);
\draw[arr] (llm) -- (tts);
\end{tikzpicture}
\caption{Cascaded voice-assistant pipeline. The streaming ASR stage and its downstream effect on LLM tool-calling accuracy.}
\label{fig:pipeline}
\end{figure*}
This work is motivated by two complementary, and at times competing, requirements for a production-grade ASR component handling financial queries: \textbf{(i) accent robustness}, since ASR trained predominantly on American-accented speech introduces systematic bias against specific customer groups rather than uniform error; and \textbf{(ii) streaming efficiency without sacrificing accuracy}, since a deployed component must emit transcripts incrementally for a responsive assistant while still supporting correct downstream tool-calling. These requirements must be optimized jointly, since the accent-robustness gap is not static -- it widens or narrows with the chosen latency operating point.

To our knowledge, no publicly available benchmark targets {\it spoken} financial queries, annotated by British-accent group, a prerequisite for measuring accent robustness in this setting rather than in general-purpose speech. We study how accent affects streaming ASR Word Error Rate (WER), and how that error propagates into end-to-end task completion, measured through the tool-calling behavior of the downstream LLM. However, WER weights all word errors equally, but effects on a financial assistant are highly unequal: errors in task-critical tokens such as merchant names, monetary amounts, dates, durations, and temporal expressions can change the resulting tool call, while errors in less informative words may have little effect on task completion. Our main contributions are: (i) CavaBench, the first internally collected, human-labeled audio benchmark for spoken financial queries, balanced across four British-accent categories; (ii) a joint evaluation of accent and acoustic robustness across various simulated acoustic conditions, linking ASR WER to downstream tool-calling Exact Match (EM).

\section{Related Work and Methods}
\label{sec:related}

\textbf{Accent and dialect bias in ASR.} Large-scale audits document systematic ASR disparities across speaker demographics: Koenecke~\emph{et al.}~\cite{koenecke2020racial} found higher WER for African American speakers across five commercial systems, and DiChristofano~\emph{et al.}~\cite{dichristofano2022global} showed ASR performance correlates with speakers' country of origin. Within British English, Markl~\cite{markl2022language} showed algorithmic bias reflects existing prejudice against non-standard varieties. Demirsahin~\emph{et al.}~\cite{demirsahin2020opensource} released open-source multi-speaker corpora of British Isles accents to support more equitable ASR. Accent benchmarks such as VCTK~\cite{yamagishi2019vctk} (British-accent read speech), EdAcc~\cite{sanabria2023edacc} (spontaneous conversational speech across global English varieties), and Common Voice~\cite{ardila2020commonvoice} (crowd-sourced multilingual speech) evaluate general accent robustness. In these, the American English WERs are consistently several points lower than British varieties -- 9.4\% vs.\ 15.2--16.6\% on VCTK and $\sim$11\% vs.\ 14--15\% on Common Voice -- an error range consistent with the realistic acoustic conditions we simulate in this work. To our knowledge, none of this work evaluates accent robustness in the financial domain: CavaBench pairs accent variability with financial queries and evaluates the full cascaded pipeline rather than ASR in isolation.

\textbf{ASR architecture families.} The evaluated systems span four families. FastConformer-transducer models~\cite{noroozi2023stateful} (Parakeet~\cite{parakeettdt2025}, Nemotron~\cite{nemotron2026}) can be configured for offline or streaming ASR depending on encoder context and decoder design. While both 0.6B models share a FastConformer backbone, Parakeet TDT pairs full-context encoder attention with a TDT decoder for high-accuracy offline decoding. In contrast, the Nemotron streaming model uses a cache-aware FastConformer encoder with causal convolutions and limited-context attention, alongside a frame-synchronous RNN-T decoder for low-latency incremental processing (Section~5). Attention encoder--decoder models (Whisper~\cite{radford2023whisper}) process fixed-length windows with full-context attention and are offline-oriented. Audio-conditioned language models (Qwen3-ASR~\cite{qwen3asr2026}) achieve strong offline accuracy but are less suited to low-latency streaming. The fourth family (Cohere Offline) is available only through a hosted API, with undisclosed architecture. RNN-T~\cite{rao2017exploring} established the streaming formulation underlying the first family. Subsequent advances in Conformer encoders~\cite{li2021better} and on-device streaming~\cite{he2019streaming} improved the accuracy--latency trade-off. However, these advances have rarely been evaluated jointly for accent robustness and downstream tool-calling correctness.

\textbf{Error propagation into downstream understanding.} Avila~\emph{et al.}~\cite{avila2023multimodal} studied how ASR errors degrade NLU in cascaded spoken language understanding, and Jung and Choi~\cite{jung2026analyzing} found that WER alone does not fully explain semantic degradation in ASR--LLM cascades. This motivates task-level metrics alongside transcription-level ones; we therefore evaluate tool-calling EM to examine how accent-related ASR errors affect downstream performance in finance. The ASR transcript is the \emph{sole} textual representation of customer intent available to the LLM, and errors often corrupt the low-frequency, high-perplexity tokens on which tool-calling depends most, such as amounts, dates and merchant names. A single substitution in such a token can silently redirect a tool call, and a user acting on the result may make a materially incorrect financial decision.

\section{CavaBench: Dataset}
\label{sec:dataset}

Cascaded ASR--LLM pipelines degrade for underrepresented accents, and in financial this can cause failed tasks, reduced user trust, and can create measurable bias. No prior benchmark targets this domain--accent intersection, motivating CavaBench ({\bf C}ascaded {\bf A}udio and {\bf V}oice {\bf A}ssistant Benchmarking). It was collected from 90 participants (50 male/40 female) via an in-house recording interface, in a clean acoustic environment, isolating accent as the primary variable. Acoustic robustness is evaluated separately, under the acoustic conditions in Section~5 (Section~\ref{ssec:acoustic-results}). Each speaker recorded at least one round of 20 queries sampled from a pool of 200 realistic financial queries, including spending summaries, budget tracking, merchant lookups, trend analysis and numerical queries. The queries were phrased naturally to elicit realistic prosody, yielding 3{,}079 utterances of which 2{,}919 were retained after quality inspection, with reference transcripts taken directly from the prompt text. Utterance duration averages 4.73\,s (range 2.16--20.82\,s; 3.94\,h total). Self-reported accent (76 raw free-text values) was grouped to support balanced group-level comparisons: Scottish ($\sim$685, $\sim$23 speakers), Southern English ($\sim$710, $\sim$23), Other British ($\sim$747, $\sim$20 -- Northern Irish, Mancunian, Yorkshire, Welsh, Midlands), Non-British ($\sim$784, $\sim$26). 

\section{Experimental Setup}
\label{sec:setup}
\textbf{ASR models and configurations.} We evaluate: (i) \emph{Nemotron Streaming} (0.6B parameters)\footnote{Fine-tuned in-house on internal data; our regulatory and data-governance constraints permit in-house fine-tuning only for open-weight models such as this one.} is evaluated using \emph{Baseline} and \emph{Finetuned} variants, at two streaming lookahead settings, LA6 and LA13 yielding four configurations.\footnote{The corpus contains various British-accented financial queries recorded under telephony-style channel and noise conditions and internally cleaned and filtered for transcription accuracy. Shorter lookahead settings were also evaluated but underperformed relative to longer lookaheads. Similarly, the larger Parakeet-RNNT-1.1B underperformed.} (ii) \emph{Parakeet-TDT-0.6B} is evaluated as both a \emph{Base Offline} variant and a \emph{Parakeet v2 0.6B} variant, the latter fine-tuned in-house on the same corpus using a gated encoder adapter on its 24-layer FastConformer encoder with TDT loss (lr=$3\times10^{-5}$). (iii) \emph{Qwen3-ASR-1.7B} is evaluated offline using  chunked pseudo-streaming mode (audio processed in 1-sec chunks), hereafter \emph{Qwen3-ASR-1.7B Chunking}. (iv) \emph{Whisper Turbo} and \emph{Distil-Whisper}, and (v) \emph{Cohere Offline} are evaluated as offline references. 
Note that in this work, we fine-tune the smaller Nemotron and Parakeet ASR models with additional in-domain data to investigate the effect of targeted adaptation on end-to-end performance.

{\bf Deployment-relevant properties.} Nemotron Streaming's cache-aware design gives it a substantial deployment edge over the alternatives: roughly 3.5$\times$ lower latency and less than half the memory footprint of Qwen3-ASR-1.7B Chunking, the closest streaming-emulating alternative, the latency of which stays high because its non-causal encoder must re-decode a growing prefix on every chunk rather than reuse cached state. None of the other evaluated models are natively streaming. {\bf LLM reasoning component.} Tool-calling is performed by \emph{Gemma-4-26B-A4B-it} reasoning model (Fig.~\ref{fig:pipeline}), prompted with the same tool schema for all ASR transcripts. {\bf Query set and gold labels.} We use the same 200-query from CavaBench (Section~4). Gold tool-call labels are obtained by running a 10-fold LLM validation on each of the 200 reference queries, and taking the majority-vote as gold, with manual inspection. {\bf Task-level metric.} Exact Match (EM) is computed over five tool-call fields: \texttt{metric}, \texttt{time\_period}, \texttt{group\_by}, \texttt{filter}, and \texttt{compare\_with}; a prediction counts as correct only if all five fields match gold. {\bf Acoustic conditions.} Each recording is evaluated under 6 groups, across all 4 accent groups: \emph{Clean} (unmodified); \emph{Noise} -- MUSAN~\cite{musan2015} additive noise/music/babble, +20 to $-5$\,dB SNR; \emph{Overlap} -- competing-speaker babble, +20 to +5\,dB SIR; \emph{Reverb} -- RIRS\_NOISES~\cite{ko2017reverb} convolution, RT60 0.3--0.8\,s; \emph{Environment} -- compound device/codec profiles with speaker-level attenuation; \emph{Telephony} -- narrowband codec/bandwidth degradation, evaluated separately. 

\section{Results}
\label{sec:results}

\subsection{Accent-robustness results (Clean condition)}
\label{ssec:clean-results}

Table~\ref{tab:wer-results} reports WER by accent group, averaged (sample-weighted) across 6 acoustic condition groups. Text normalization lowercases text, strips punctuation, and canonicalizes numbers and currency. First, fine-tuning helps Nemotron Streaming without widening the accent gap: Nemotron (FT-LA13) improves overall WER from 18.2\% to 16.1\%, and narrows the max-min accent spread (1.5\% vs.\ 2.3\%, the fairness metric in Section~7). Second, applying the same procedure to Parakeet narrows the accent spread from 2.7\% to 1.7\%, with only a marginal improvement in overall WER (16.6\% to 16.5\%) and EM is not improved. Third, accent sensitivity is model-dependent: Qwen3-ASR-1.7B Chunking has the largest spread of any model (9.4\%, driven by Scottish speech collapse at 22.2\%). Cohere Offline achieves the third-lowest overall WER (15.3\%) and one of the tightest accent spreads (1.6\%). In the Clean-only condition, overall EM ranges from 75.1\% (Nemotron Base-LA6) to 80.7\% (Qwen3-ASR-1.7B Offline); with fine-tuning raising Nemotron Streaming's (LA13) EM from 75.5\% to 78.2\%. 

\begin{table}[t]
\centering
\caption{WER (\%) by accent group, averaged over all 6 acoustic condition groups, with each model's overall average WER. \textbf{Bold} marks the top-3 (lowest-WER) models.}
\label{tab:wer-results}
\footnotesize
\setlength{\tabcolsep}{3.5pt}
\begin{tabular}{@{}lccccc@{}}
\toprule
\textbf{Model} & \textbf{Non-Br.} & \textbf{Other Br.} & \textbf{Scot.} & \textbf{S.\ Eng.} & \textbf{Avg} \\
\midrule
Nemotron (Base-LA13) & 18.1 & 19.7 & 17.4 & 17.4 & 18.2 \\
Nemotron (FT-LA13) & 16.6 & 16.7 & \textbf{15.2} & \textbf{15.8} & 16.1 \\
Nemotron (Base-LA6) & 19.5 & 22.1 & 19.4 & 18.8 & 20.0 \\
Nemotron (FT-LA6) & 17.9 & 18.9 & 16.7 & 17.0 & 17.6 \\
Qwen3-ASR-1.7B Offline & \textbf{10.2} & \textbf{11.1} & \textbf{11.4} & \textbf{11.8} & \textbf{11.1} \\
Qwen3-ASR-1.7B Chunking & \textbf{12.8} & \textbf{13.3} & \underline{22.2} & \textbf{12.9} & \textbf{15.2} \\
Parakeet FT Offline & 16.6 & 17.1 & 15.4 & 16.7 & 16.5 \\
Parakeet Base Offline & 16.2 & 18.2 & 15.5 & 16.7 & 16.6 \\
Whisper Turbo & 14.9 & 17.8 & 16.1 & 18.2 & 16.7 \\
Distil-Whisper & 16.4 & 18.8 & 16.7 & 17.7 & 17.4 \\
Cohere Offline & \textbf{14.5} & \textbf{16.1} & \textbf{14.8} & \textbf{15.8} & \textbf{15.3} \\
\bottomrule
\end{tabular}
\end{table}



\subsection{Robustness across acoustic conditions}
\label{ssec:acoustic-results}

\begin{figure}[t]                                   
\centering
\includegraphics[width=\columnwidth]{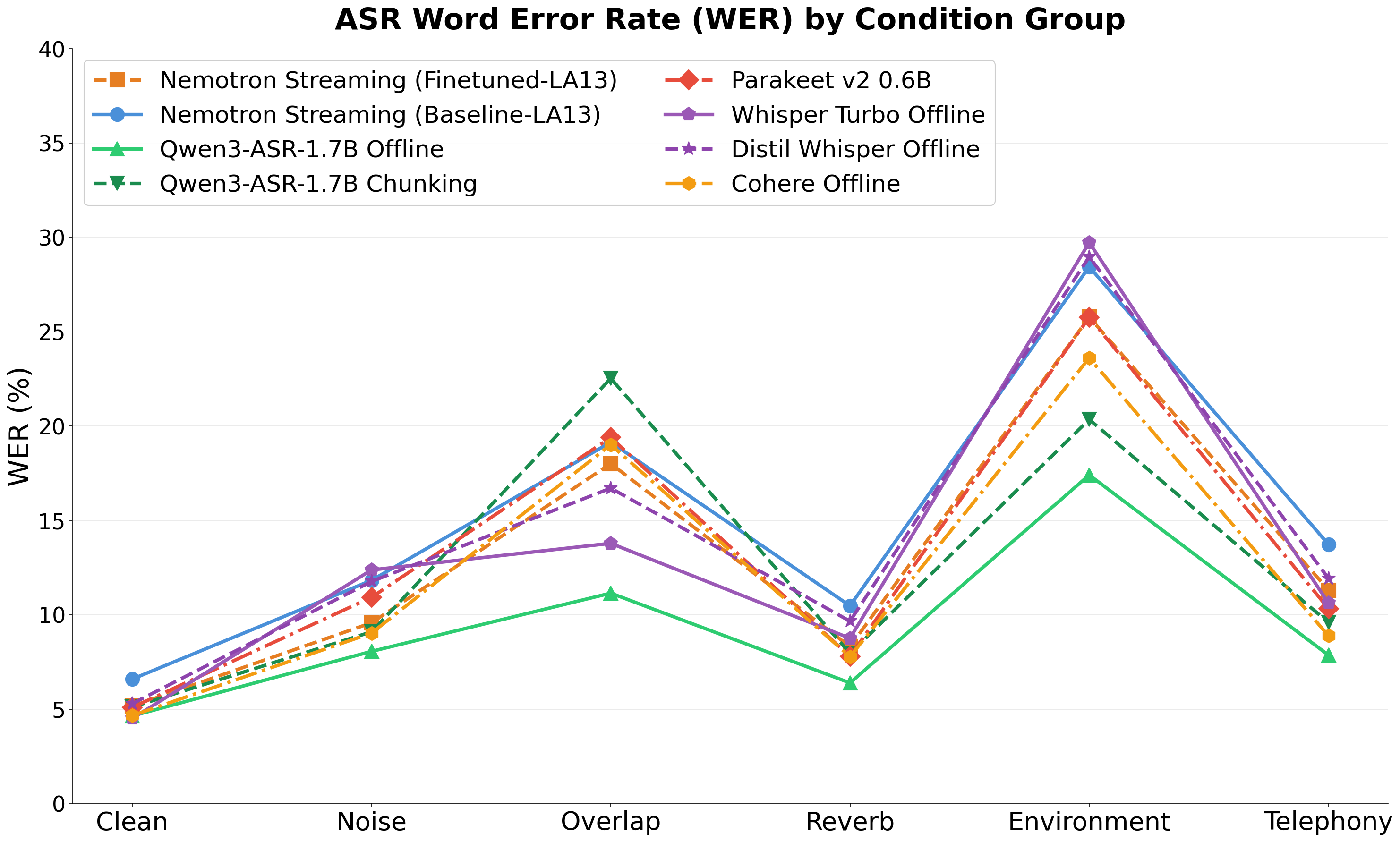}
\caption{Word Error Rate (\%) by condition group, per model, averaged over the 4 CavaBench accent groups.}
\label{fig:wer-grouped}

\vspace{2mm}                                        
\includegraphics[width=\columnwidth]{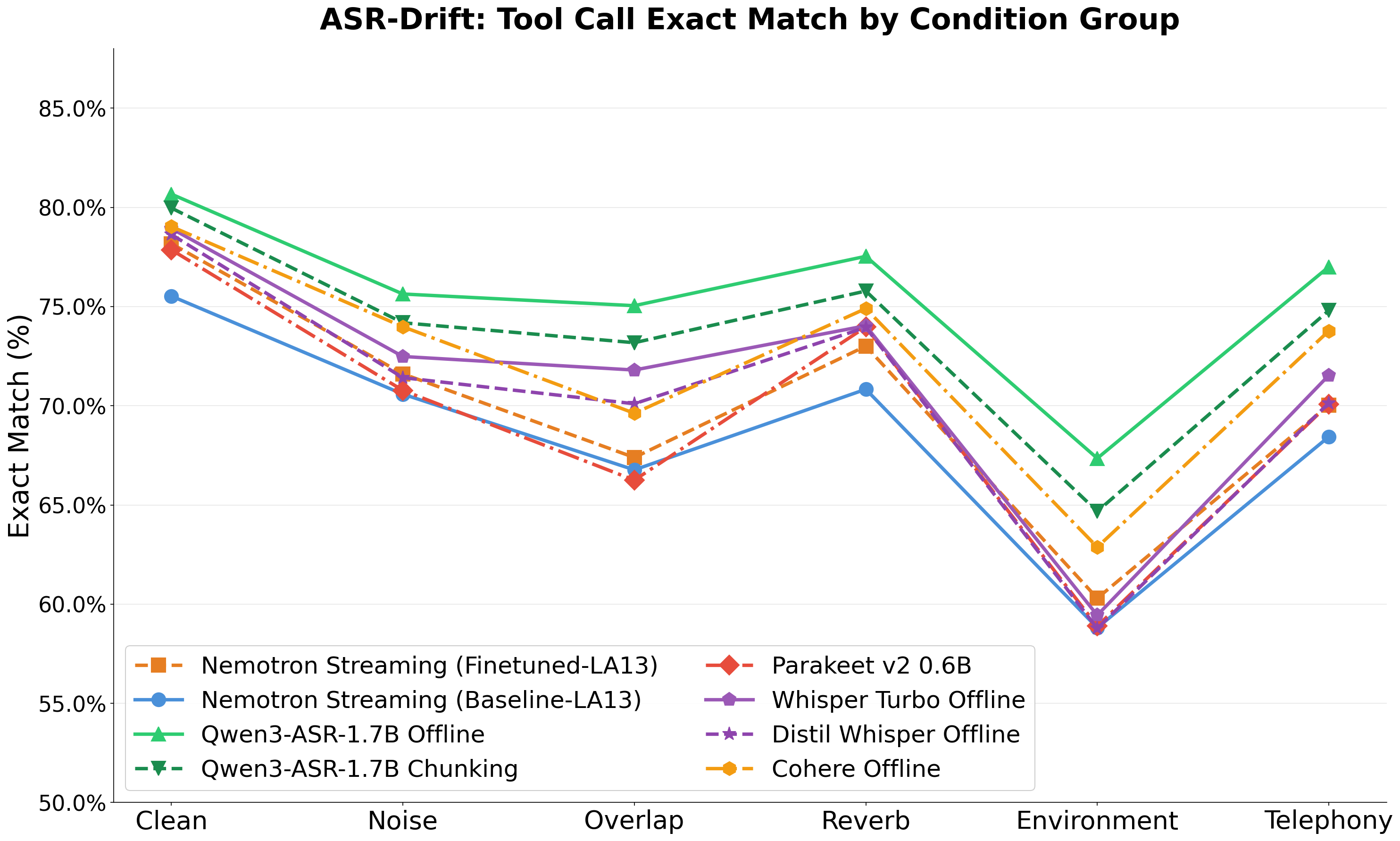}
\caption{Tool-calling Exact Match (\%) by condition group, per model, averaged over the 4 CavaBench accent groups.}
\label{fig:em-grouped}
\end{figure}

ASR robustness decreases monotonically across conditions: Clean (6.0\% WER) $<$ Reverb (9.8\%) $<$ Noise (11.8\%) $<$ Telephony (12.9\%) $<$ Overlap (20.0\%) $<$ Environment (28.6\%) (Figs.~\ref{fig:wer-grouped}--\ref{fig:em-grouped}), confirming joint degradation (Environment) is substantially harder. All models do not degrade uniformly: Whisper Turbo, differing from its distilled variant only in the decoder size, has the steepest single-condition collapse (29\% WER at Environment), while Qwen3-ASR-1.7B Offline is the most robust throughout, retaining the highest EM except for Reverb. Cohere Offline is behind only the two Qwen3-ASR-1.7B variants. The Nemotron Streaming baseline, comparable in scale to Parakeet but 2-4$\times$ smaller than the Whisper and Qwen3-ASR-1.7B shows less robustness, likely because it uses less acoustically diverse pretraining data, and model size. The fine-tuned models reduce the gap at Telephony, matching the fine-tuning corpus's own domain.

Averaged over the acoustic conditions, overall WER and EM are strongly anti-correlated across models ($r=-0.93$), indicating that WER is generally informative for model-selection. However the relationship varies by accent: correlations range from $r=-0.87$ to $-0.98$ for 3/4 groups but weaken substantially for Scottish ($r=-0.29$), largely driven by Qwen3-ASR-1.7B Chunking's disproportionate Scottish WER spike. All four Nemotron Streaming variants show strong within-model correlations ($r=-0.74$ to $-0.96$) and the same ranking pattern -- Southern English achieves marginally higher EM than Scottish despite slightly higher WER -- indicating that the pattern is consistent across fine-tuning and lookahead settings. Comparing LA6 (shorter context, lower latency) against LA13 (longer context, higher latency) isolates the lookahead effect. The Baseline configuration improves from 20.0\%/64.6\% (LA6) to 18.2\%/66.5\% (LA13), while the Finetuned configuration improves from 17.6\%/66.4\% to 16.1\%/67.9\%. This corresponds to a consistent but \textit{modest} improvement of 1.5--1.8\% in WER and 1.5--1.9\% in EM. 



\subsection{Error analysis}
\label{ssec:error-analysis}

\textbf{Impact on downstream tool-calling.}
In the Clean condition, 21.9\% of tool-call predictions were incorrect: \texttt{time\_period} is the largest failure mode (33.5\%), followed by \texttt{compare\_with} (25.4\%) and \texttt{group\_by} (25.3\%); abstention accounts for 16.2\%, \texttt{filter} for 12.3\%, false positives for 8.7\%, and \texttt{metric} is rarely wrong (0.7\%) -- numeric/temporal arguments drive downstream failure, not metric choice (Section~3.2). This interacts with accent in ways Table~\ref{tab:wer-results}'s condition-averaged spread cannot capture. {\it Other British} bundles several distinct regional accents into one category, yet it is the lowest-EM group under most individual conditions for every evaluated model, not just on average. The harder robustness challenge is accent diversity itself rather than any single accent; and fine-tuning narrows this gap.

\textbf{Task-critical token preservation matters more than aggregate WER.} A common UK retailer name is transcribed correctly in only 56.3\% of 2{,}703 utterances (197 distinct variants; 24.4\% dropped outright), and ``last quarter'' survives verbatim in just 43.3\% of 61{,}786 transcriptions (36.5\% dropped, 19.4\% flattened to a bare ``Q1''--``Q4'' label, occasionally ``last water''); day counts fare better (85.9--98.4\% correct) but show a systematic 60$\to$``sixteen''/``six'' confusion -- unlike the abstentions below, this substitution returns a confident, wrong answer (a 60-day spending query silently computed over 6 or 16 days instead). These are exactly the tokens tool-calling depends on.

\textbf{Qualitative cases.} Manual inspection reveals two representative failures: an Other British-accent query, \emph{``How much have I spent at Tesco last year?''}, was transcribed as \emph{``Hi Monte Santesco last year''} -- the merchant name was destroyed, triggering abstention; and \emph{``List purchases over £600 in the last 90 days''} became \emph{``This purchase is over six hundred pounds in the last ninety days''}, the dropped imperative again causing abstention. In both cases, a lower-WER system preserved the reference verbatim on the same audio; for the customer, abstention at least surfaces as a retryable stall, unlike the silent wrong answer above.

\textbf{WER is necessary but not sufficient.} Qwen3-ASR-1.7B Chunking has the worst WER observed (49.8\% at overlap +5\,dB) yet retains 67\% EM there, since its errors are mostly filler padding while task-critical financial keywords survive at a higher rate than in lower-WER competitors (e.g., ``groceries'' 86.8\% vs.\ 76.3\% for Whisper Turbo). Consistently, Nemotron Streaming (Finetuned-LA13) beats both Parakeet variants on WER and EM despite streaming (16.1\%/67.9\% vs.\ 16.5--16.6\%/67.1--67.8\%), and Cohere Offline -- with no fine-tuning at all -- still edges past it (15.3\%/70.1\%), though both trail Qwen3-ASR-1.7B Offline (11.1\%/73.7\%). \emph{Which} words are garbled can matter more than \emph{how many}: preserving task-critical keywords and intent-bearing tokens is more directly tied to correct tool-calling and end-to-end user experience than aggregate WER alone.



\section{Conclusion and Future Work}
\label{sec:conclusion}
We conducted a study using CavaBench, our internally collected financial-query benchmark, to identify general trends in how accent, model architecture, and acoustic conditions jointly shape end-to-end ASR accuracy. WER and Exact Match are strongly correlated overall, but different accents reveal different challenges: a single model's WER spike breaks the correlation for Scottish, while for Other British -- which bundles several distinct regional accents into one category -- accent diversity itself, not any single accent, is the harder challenge. Aggregate WER also masks the unequal cost of errors: preserving task-critical tokens (merchant names, amounts, dates) matters more than reducing total word errors. The robustness gap itself traces to training data rather than architecture: targeted fine-tuning closed much of it for the smaller streaming model, pointing to efficient streaming models trained on more diverse data as a path to closing the gap with offline counterparts.

\end{document}